# Towards skin-acetone monitors with selective sensitivity: dynamics of PANI-CA films

Anthony Annerino[1,¶,*], Michael Faltas[1,¶], Manoj Srinivasan[2,3], and Pelagia-Irene Gouma[1,2]

[1]Material Science and Engineering, The Ohio State University, Columbus, OH 43210, USA

[2]Mechanical and Aerospace Engineering, The Ohio State University, Columbus, OH 43210, USA

[3]Program in Biophysics, The Ohio State University, Columbus, OH 43210, USA

[*]Corresponding author

E-mail: Annerino.2@osu.edu (AA)

[¶]These authors contributed equally to this work.

# Abstract

Most research aimed at measuring biomarkers on the skin is only concerned with sensing chemicals in sweat using electrical signals, but these methods are not truly non-invasive nor non-intrusive because they require substantial amounts of sweat to get a reading. This project aims to create a truly non-invasive wearable sensor that continuously detects the gaseous acetone (a biomarker related to metabolic disorders) that ambiently comes out of the skin. Composite films of polyaniline and cellulose acetate, exhibiting chemo-mechanical actuation upon exposure to gaseous acetone, were tested in the headspaces above multiple solutions containing acetone, ethanol, and water to gauge response sensitivity, selectivity, and repeatability. The bending of the films in response to exposures to these environments was tracked by an automatic video processing code, which was found to out-perform an off-the-shelf deep neural network-based tracker. Using principal component analysis, we showed that the film bending is low dimensional with over 90% of the shape changes being captured with just two parameters. We constructed forward models to predict shape changes from the known exposure history and found that a linear model can explain 40% of the observed variance in film tip angle changes. We constructed inverse models, going from third order fits of shape changes to acetone concentrations where about 45% of the acetone variation and about 30% of ethanol variation are captured by linear models, and non-linear models did not perform substantially better. This suggests there is sufficient sensitivity and inherent selectivity of the films. These models, however, provide evidence for substantial hysteretic or long-time-scale responses of the PANI films, seemingly due to the presence of water. Further experiments will allow more accurate discrimination of unknown exposure environments. Nevertheless, the sensor will operate with high selectivity in low sweat body locations, like behind the ear or on the nails.

# Introduction

The presence and concentrations of gaseous products of metabolic processes in the breath, often volatile organic compounds (VOCs), are well-established indicators of the state of one's health, and measuring these has been the subject of much research [1–12]. However, even the apparently trivial work of blowing into a breathalyzer is a task that requires active intent and thus may suffer from subject compliance issues for continuous health monitoring. Further, the results of such breathalyzer analyses provide only a momentary snapshot. This has given rise to a new, totally non- intrusive vision for health monitoring. This new vision, the subject of many recent investigations, is that of continuous biomarker monitoring via wearable devices [13–25].

Most investigations aimed at monitoring biomarkers with wearables, including the few fully functional and available devices, involve analysis of sweat [13,15,16,18,18,25–29]. However, sweat-based continuous monitoring is challenging as the contents of sweat can change independently of the health conditions being monitored. This shortcoming is best put by Ganguly et al. who acknowledge that the "influence of a sweat constituent, such as an electrolyte, on a given biomarker (in the sweat) and the corresponding sensor readout is not accounted for," by the currently available sweat monitors because standardizing sensor readouts across the multitudinous compositions of sweat is such a monumental challenge [13]. Further, sweat is not continuously emitted in stable quantities, again making continuous monitoring challenging.

One proposed solution for truly continuous, truly non-invasive monitoring is skin gas sensing: that is, via analysis of the gaseous biomarkers continuously and ambiently emitted directly from the skin. Biomarkers observed in the breath are known to be emitted from the skin, continuously and in quantities proportional to quantities seen in the breath, allowing for easy selection of target analytes [4–6,30–33]. Further, the continuous emanation of skin gases means that their analysis could proceed non-invasively and with no conscious effort from the subject. Thus, skin-gas biomarker monitoring combines the most attractive features of breath sensing and sweat sensing.

The first step toward novel wearable devices for continuous, non-invasive skin-gas sensing is the establishment of sensing materials that are sensitive and selective to analytes of interest. One such analyte of interest that is found in breath and skin-gas is acetone, the concentration of which in the breath has been studied in relation to blood sugar levels and fat burning rates and could shed light on the progression of diabetic ketosis and weight loss [2,4,6,8,11,28,34]. In this manuscript, we examine the sensitivity and selectivity of a novel chemo-mechanical actuator, developed to detect gaseous acetone [14].

The chemo-mechanical actuator we propose is a polyaniline-cellulose acetate (PANI-CA) composite film that bends upon exposure to gaseous acetone; we refer to this response as being chemo-mechanical, as the film responds with mechanical deformation when exposed to acetone [14]. This chemomechanical response is to be contrasted from electrochemical responses previously used for acetone sensing [35]. Previous tests of this novel chemo-mechanical actuating material focused on the responses of single exposures to dilutions of acetone in water and dilutions of three alcohols in water [14]. The alcohol tested in this previous investigation that elicited the greatest bending response was ethanol, which is not just an important interferent to be aware of but also a target analyte not just for measuring intoxication but also liver disease [12]. This previous investigation of the novel PANI-CA chemo-mechanical actuator and the relevance of acetone and ethanol inspired our current investigation of the PANI-CA response to repeated exposure to various dilutions of acetone and ethanol.

Polyaniline (PANI), a conjugated polymer derived from the organic solvent aniline, has long been known to exhibit intrinsic electrical conductivity. PANI can exist as fully oxidized pernigraniline, half oxidized-half reduced emeraldine, or fully reduced leucoemeraldine, and each of these oxidation states has a fully protonated salt and a fully deprotonated base version; the emeraldine salt is known to conduct most strongly [1]. The structure of the leucoemeraldine salt is presented in Fig 1a. Cellulose acetate (CA) is a derivative of the enormously abundant and inexpensive biopolymer cellulose. Because CA has much more robust mechanical properties than PANI, PANI and CA have been combined as a composite, PANI-CA (Fig 1b), to take advantage of PANI's conductivity while mitigating PANI's relatively poor stability and mechanical properties [2].

**a) Leucoemeraldine Salt Polyaniline Structure**

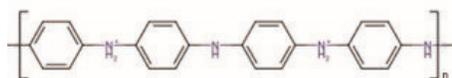

**b) A Representative PANI-CA Film**

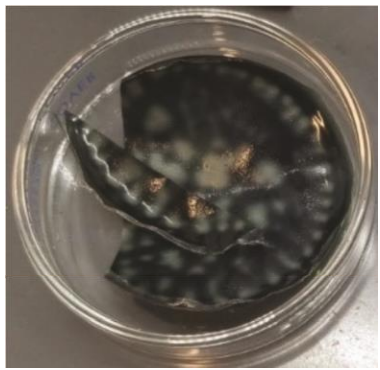

**c) An ongoing headspace exposure test**

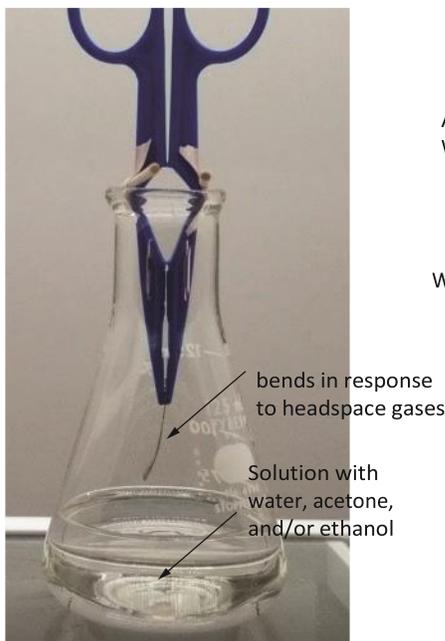

bends in response to headspace gases

Solution with water, acetone, and/or ethanol

PANI-CA strip

**d) Solutions considered**

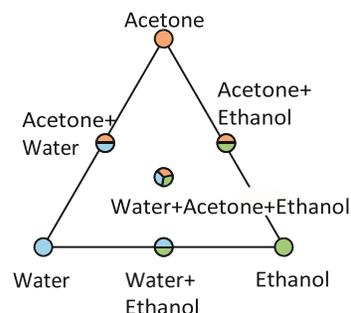

**Fig 1. A Novel Chemo-mechanical Actuator.** a) The structure of leucoemeraldine salt of PANI. b) A film of polyaniline cellulose acetate (PANI-CA) composite. c) A PANI-CA film bending in response to gases in the headspace of the conical flask. The solution in the flask has water, acetone, and/or ethanol.

This PANI-CA composite has previously been shown to exhibit chemo-mechanical actuation upon exposure to gaseous acetone, bending in response to acetone exposure. Presented here is a study of the repeated exposure of strips cut from PANI-CA composite films to the headspaces above samples of pure acetone, pure ethanol, and mixtures of these with each other and with water [2]. Our study is aimed at establishing the sensitivity, selectivity, and reproducibility of sensing acetone vapor by measuring the chemo-mechanical actuation of the PANI-CA composite films.

# Methods

## Fabricating the PANI-CA films

The novel, chemo-mechanical actuating films were prepared by first proton-doping leucoemeraldine base PANI with concentrated hydrochloric acid for 24 hours, then washing off the acid with water and drying at room temperature in an open container. Dry, doped PANI was suspended in acetone and combined with 50,000 g/mol number-average molecular weight CA in a 1:5 mass ratio. This mixture was then dried in a loosely covered container at room temperature to leave behind a composite film. Fig 1b displays an image of a PANI-CA composite film representative of those used in the experiments presented here.

## Testing the chemo-sensitivity of PANI-CA films

Strips measuring approximately 6 mm by 25 mm were cut from full PANI-CA films. We tested 8 strips, obtained by cutting out 2 strips each from 4 larger films. Any small residual curvature in these strips was left uncorrected before testing. Headspace exposure environments were prepared by pouring 60 mL of various pure solvents or mixtures of solvents into 125 mL conical flasks. The headspace exposure environments used here were those above the solvents and mixtures that are identified in Table 1. Each PANI-CA strip was exposed to each of the 6 headspace environments 5 times, giving a total of 30 exposure tests with each PANI-CA strip. The exposure order for each PANI-CA strip was randomized (the random order drawn from a uniform distribution over all permutations using MATLAB); 5 unique sets of random permutations of the integers 1 through 6 were generated and strung together to make a string of 30 exposure tests for each test specimen.

**Table 1. Volatile Mixtures Tested.**

| Solution ID | Solution Identity |
|---|---|
| 1 | 60 mL acetone |
| 2 | 60 mL ethanol |
| 3 | 30 mL acetone+30 mL ethanol |
| 4 | 30 mL acetone+30 mL water |
| 5 | 30 mL ethanol+30 mL water |
| 6 | 20 mL acetone+20 mL ethanol+20 mL water |

Each headspace exposure test consisted of recording a video of the PANI-CA strip suspended in the headspace above a solution, as shown in Fig 1c. All headspace exposure tests of a given PANI-CA strip were conducted in immediate succession with 60 seconds between exposure tests. In the 60 seconds between successive exposure tests, the PANI-CA strip was suspended in an empty 125 mL conical flask as if a headspace exposure test were ongoing. The duration 60 seconds was selected to allow for substantial recovery of the PANI-CA strip from its final bent state resulting from the previous test, based on time-constants of deformation observed in pilot trials. Flasks not being actively used for a test were covered with paraffin film to minimize changes in headspace composition between tests. Each headspace exposure test was run until the PANI-CA strip reached a clear maximum deflection or appeared to reach a steady state of deflection, as determined by visual inspection. All headspace exposure tests were recorded with a Samsung Galaxy S8+ smartphone mounted in a tripod. The 60 second interludes between exposure tests were not recorded.

## Computer vision-based automatic tracking of filament shape

First, the videos were analyzed by manually determining the timestamp of the maximum deflection in each test video. After this manual analysis, we performed computer vision-based tracking of the bending film, described in sequence below. For this computer-based analysis and for the rest of the manuscript, only 4 out of the 8 strips were analyzed: specimens 1 through 4, collected from films 1 and 2, exhibited high degrees of non-ideal deformation, twisting laterally out of the video image plane. We ignored films exhibiting such out of plane deformation in this study; we will examine them in a future study or engineer forming processes to prevent them in future work.

Videos were first cropped and magnified to focus on the filament, cropping out inconsequential visual details such as the flask holding the solution. These cropped videos were then processed by a bespoke MATLAB program, which first splits the video into frames at 30 Hz, and then converts them into binary black and white images (Fig 2). From the binary black and white image, we extracted the shape of the filament as follows: we computed a single contiguous path from the root of the filament to the tip of the filament by scanning each horizontal line of pixels, and then computing the mean of the black pixels on that line as the midpoint of the filament. This procedure works well when the filament monotonically goes down from root to tip and does not curl upwards. To also track the midpoint when the curve curls upwards, for each horizontal like of pixels we also performed a k-means clustering, with two clusters. If the two clusters are substantial and are separated by the typical thickness of the filament, then it means that the current level has two sections of the filament, one going down and another going up after having curled up. We then stepped from the root downward following either the global mean or one of the two clusters of the k-means process and rearranging these points in a list in a manner that they form a single path that went from root to tip even when the filament curled up. We defined the filament tip as the last 25% of the total length along the filament and computed the angle that this tip made with the horizontal. This mean tip angle was computed by considering all the points that comprise the tip (last 25%), performing a principal components analysis (PCA), and then picking the direction of the first principal component vector. By definition, this is the direction in which the cluster of points have the greatest variance, which we define to be the filament direction.

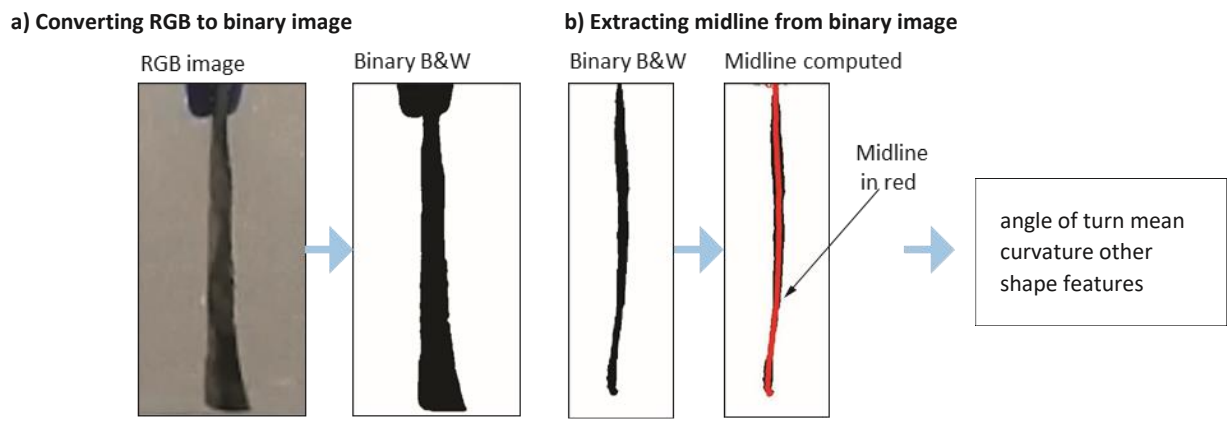

**Fig 2. Image Processing and Computer Vision.** Some steps in the automatic analysis of the video are outlines. a) RGB images from the video frames are converted to binary black and while images. b) Binary images are processed to extract a contiguous midline through the filament. This midline path is then used to determine shape features such as angle of turn, curvature, etc.

## Tracking the filament with Manual labeling and DeepLabCut

We used two other methods of tracking the filament shape. First, we did manual tracking of the filament shape by the following method: a computer program displayed a frame of the video at 5 seconds apart to a human user; the human user then clicks on 7 points on the filament, starting from root to tip, and then moving to the next frame. The 7 points are asked to be roughly equidistant, but for our analysis, they need not be exactly equidistant. We used cubic spline interpolation to densely fill the rest of the filament between these points. This manual tracking provided a 'ground truth' for evaluating the accuracy of the automatic tracking. We performed this manual tracking for two of the filaments.

Our automatic tracking algorithm from the previous paragraph relied on relatively classical image processing and machine learning methods: binary thresholding and k-means clustering. We sought to test whether more modern computer vision techniques could track the filament better. To do this, we used the Deep Neural Networks based software called DeepLabCut [36]. This software allows a user to train the system by clicking on a few key points on a few frames (5-20 frames, say), and then using 'transfer learning' to allow it to deduce the salient features of these key points so that it may identify them in the rest of the frames [37]. We processed one of the longer videos with this procedure.

## Filament shape complexity: Dimensionality reduction

While the filament shapes look relatively simple to the human eye, we sought to see if the set of filament shapes was indeed as low dimensional as it seemed. So, we performed principal components analysis on the full set of points comprising the tracked filament shapes, across all time and across all exposure conditions. This PCA then results in the principal components, which are the directions along which the shape variance is highest (first PC), second highest (second PC) and so on. This method allows us to see what the effective dimensionality of the filament shape is — that is, how many numbers it takes to represent the shape of the filament to a certain degree of accuracy.

## Forward and inverse predictive models

We constructed two types of models: (1) forward models that attempt to predict the filament bending response based on acetone and ethanol exposures, and (2) inverse models that attempt to predict the acetone and ethanol concentrations based on characteristics of the filament bending response.

### Forward models

As an example of a forward model, we computed a linear model from acetone and ethanol fractions in solution to the delta change in the tip angle over a trial as well as the dominant time constants of the bending response. We tested whether inclusion of the initial angle for each trial in the inputs changed the predictive power. A more general forward model would be a time-series model or differential equation model that aims to explain the shape changes of the filament given the changes in the acetone or ethanol fractions.

### Inverse models

As an example of inverse models that go from the bending response to the acetone and ethanol concentrations, we first fit a triple exponential to the tip angle transient response to every new exposure: $\vartheta(t) = a_0 + a_1 e^{-\lambda_1 t} + a_2 e^{-\lambda_2 t} + a_3 e^{-\lambda_3 t}$, with the coefficients ($a_0$, $a_1$, $a_2$, $a_3$) and the exponents ($\lambda_1$, $\lambda_2$, $\lambda_3$) being the unknowns solved by fitting to the data. These parameters describe how the tip angle changes. We then fit a linear model that took these parameters as input during each exposure phase to predict the corresponding acetone and ethanol concentrations.

# Results

## Validating the automatic tracking of filament shape

In Fig 3a, we can see a depiction of the filament shape changes, specifically, its dramatic bending as it is exposed to acetone. Such automatic tracking of the filament shape was well-correlated with the manually collected tracking data. Specifically, as seen in Fig 3b, the tip angle variation computed using automatic tracking is closely predicted by the tip angle changes computed by manually obtained tracking data, having a 93% $R^2$ for a cubic polynomial model that predicts the automatic results from the manual results.

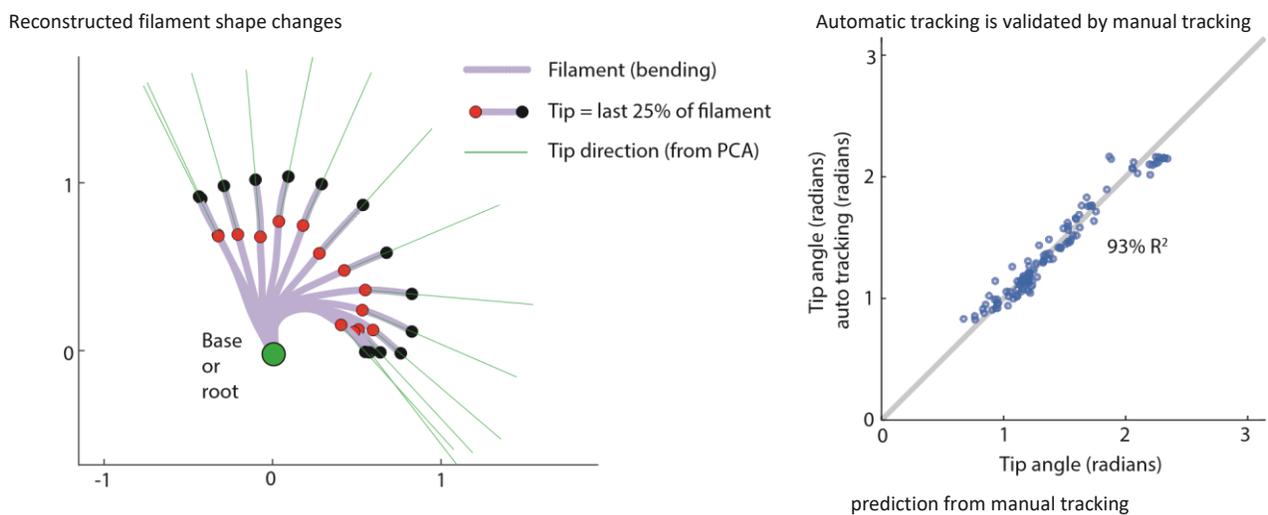

**Fig 3. Validating Automatic Filament Tracking.** a) Automatically reconstructed filament shape changes are shown for one trial. The final 25% of the filament length, labeled by the red and black solid circles are used to compute the tip direction (green). b) Automatic computer-vision-based tracking the filament agrees well with manual tracking of the filament.

## Deep neural network based filament tracking using DeepLabCut

As seen in Fig 4, the automatic computer vision-based tracking of 7 key points on the filament was not as good as the classical image processing based approach, especially in the initial few frames of the video. The initial few frames of the video were critical for the tasks herein, because those frames represented the initial few seconds of exposure, which often involved fast changes in filament shape. Because this approach was not clearly better than our other approaches, we did not pursue this approach further for this project. The software, DeepLabCut, may have had difficulty with tracking the 7 key points because, except for the root and the tip, the points do not lie on visually distinctive regions.

Computer vision tool DeepLabCut used to track the filament. Early frames have more error

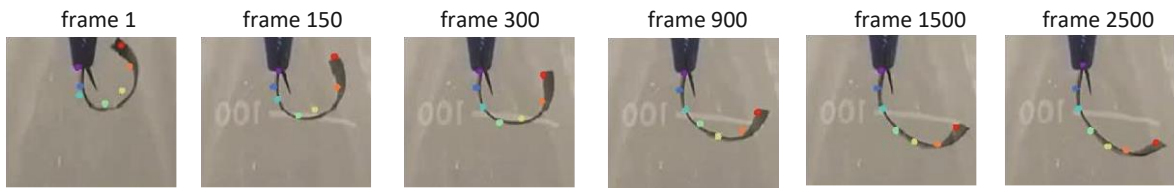

frame 1     frame 150     frame 300     frame 900     frame 1500     frame 2500

**Fig 4. Tracking with deep neural networks.** DNN-based tool DeepLabCut produced tracking well correlated with the manually obtained ground truth. The colored dots are the key point location predicted by the tool. The predicted key points had greater errors in the initial frames of the video compared to the later frames.

## Simplicity and dimensionality of film shape

We found that the filament shape can be quite accurately represented using 1 to 4 parameters. Specifically, Fig 5a shows the cumulative variance explained by the first few principal components (PCs). The first PC accounts for around 80% of the variance, and the first two PCs can capture 95% of the shape variance. This means that representing the shape of the filament as a linear combination of the first two PCs (plus the mean shape) can represent the filament with high accuracy. As seen in Fig 5b, a linear combination of the first two PCs essentially changes the overall orientation of the filament. Using the third and the fourth principal component in the linear sum results in capturing over 99.5% of the variance. Given the high explanatory power of a few principal components, we posit that describing the filament shape with one or two variables would be sufficient to build predictive models between filament shape and volatile compound concentrations.

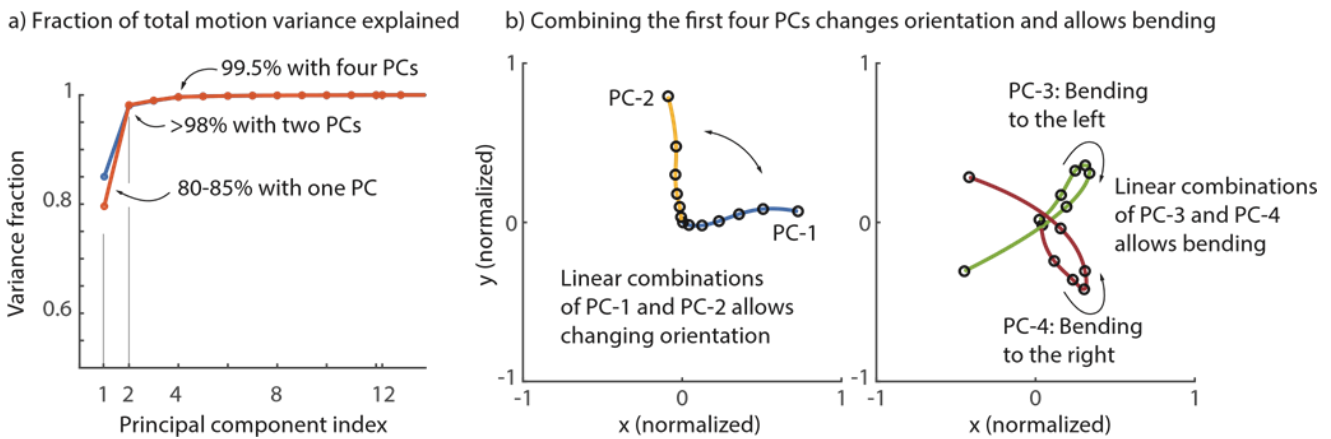

**Fig 5. Dimensionality of the Bending Filament.** a) We find that one principal component captures about 80% of the variance across a whole thirty-exposure trial. Two principal components capture over 95% and four capture over 99.5%. Curves are shown for specimen 5 (blue) and specimen 6 (red). b) The shapes of the principal components are shown; linear combinations of the first two PCs produce different orientations and the third and the fourth PCs control bending to the left versus the right.

## Acetone and ethanol concentrations affect the speed but not steady state of the filament

We did not see any systematic correlation between the steady state bending state (tip angle) of the filament and the acetone/ethanol concentrations. However, we do see systematic changes in the speed or vigor of the response with the nature of the solution (Fig 6). A loose trend can be seen here that **with increasing concentrations of acetone, a faster response time is seen**; here the response time reported is the time to maximum deflection.

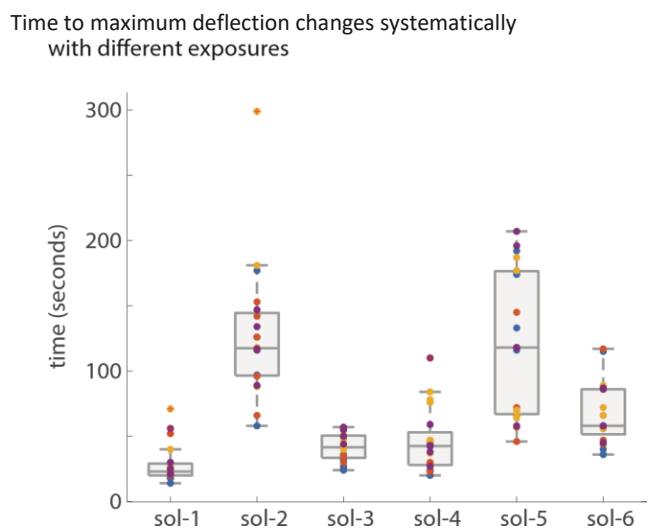

**Fig 6. Speed of Deflection.** The time to maximum deflection changes systematically between exposures to the different acetone and ethanol concentrations but there is a large range in the responses.

## Forward and inverse predictive models

Figs 7a and 7b show continuous time-series of acetone fraction, ethanol fraction, and the corresponding tip angle changes over nearly 6000 seconds. The goal of predictive models is to relate the volatile compound concentrations to the tip angle (Figs 7a and 7b), and vice versa. We found that going from the shape changes (described via parameters of a triple exponential) to the acetone concentration via a linear model has a 45% $R^2$ (which is 67% correlation between prediction and actual value). This $R^2$ value indicates some predictive ability for the shape changes, but with some errors. Indeed, only the constant term and the exponents of the exponential matter for this linear model; including the other coefficients increases $R^2$ by only 5%. Conversely, building a forward model to predict the change in tip angle from the acetone and ethanol concentrations via a linear-model results in similar explanatory power (about 40% $R^2$). We posit that a key mechanism for this not-too-high explanatory power is the presence of history dependence (hysteresis), which could be captured by models with additional state variables – for instance, characterizing the internal state of the filament and thus its propensity to bend.

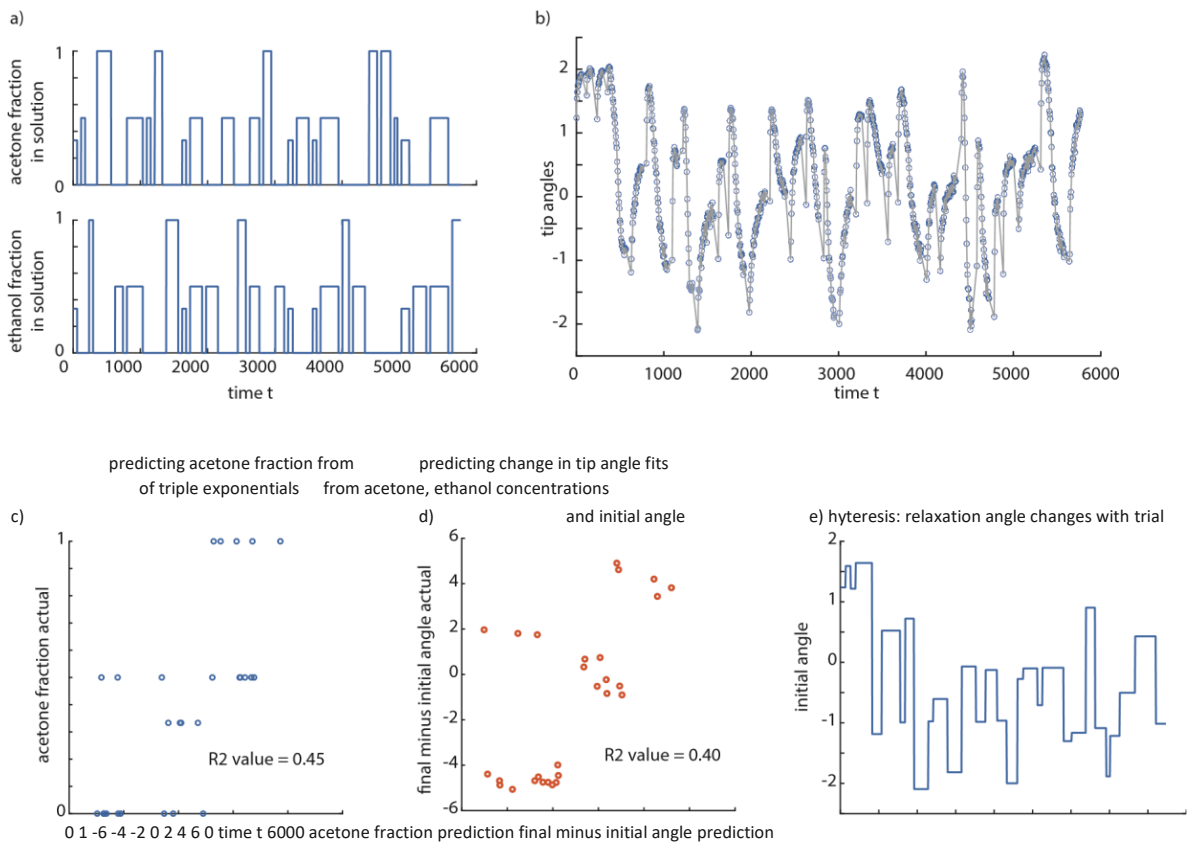

**Fig 7. Predictions from Filament Bending and Predictions of Filament Bending.** a) A sample series of 30 exposures to varying solution fractions of acetone, ethanol and water are shown. b) The filament tip angles (relative to some absolute orientation) are shown as a time series. These correspond to the exposures in panel-a. c) Predicting the acetone fraction in solution using a linear model based on fitting a triple exponential to transients gives about 0.45 $R^2$ value, suggesting some but not perfect predictive ability. d) Predicting the change in tip angle over a trial using the acetone fraction, ethanol fraction, and the initial angle using a linear model gives about 0.4 $R^2$ value. e) The initial angle for each trial – which was the angle to which the filament relaxed after the previous trial – changes over time, suggesting hysteresis, or slower time-scale processes over many tens of minutes.

## Discussion

The most striking result of this work is the degree of exposure-history dependence that the samples displayed. The next most significant result reported here is the trend in response time reported in Fig 6. First there is the difference in response time seen between exposures to solutions 1 and 2, which were pure acetone and pure ethanol. Then, there is the similarity in response time seen between most exposures to solutions 3 and 4, which both consist of half acetone by volume fraction but are alternatively completed by ethanol and water. Together, **these demonstrate that not only is the sensitivity to acetone greater than the sensitivity to ethanol when considered as pure solvents** but also that **the presence of acetone overpowers the response to ethanol** so much that responses to solutions 3 and 4 were largely indistinguishable. This is further supported by the response times to exposures to solution 5, which was composed of equal volume fractions of ethanol and water; such response times mostly resembled the response times to exposure to pure ethanol. This raises the question of the effect of water on the measurements, which in turn guides us to position the sensor strips behind the ears or on the nails, for example.

Other important result includes the fact that our original tracking program strongly agreed with the manual tracking and the fact that nearly all of the bending variance can be attributed to 4 principal components. These mean that no new computer vision tracking program will have to be written for future experiments and that the performance consistency of samples made in the future can be measured against this benchmark of 4 principal components to totally describe their motion. Furthermore, it has been made apparent that additional sample modification is required for the use of DNN tracking to monitor sample bending; some method of making points on the samples more visually distinctive would be needed. An ideal marking method would involve making marks that are easy for the computer to see but also do not influence the motion of the sample by mechanical or chemical means.

Two other significant questions are raised here that will be the focus of future investigation. The first of these is if the tremendous magnitude of exposure history dependence seen here retains this level of influence when the entire history of the sample includes only exposures to analyte concentrations on the much lower magnitudes representative of VOC concentrations that are emitted from human skin. The second is what extra steps can be taken to produce samples that behave more uniformly therefore efforts are underway to standardize the size, shape and thickness of the sensing strips.

## Acknowledgements

This study was supported by NSF SCH-INT grant number IIS 2014506 awarded to PG and MS. The authors declare that they have no competing interests.